# AN INTEGRATED RANKING ALGORITHM FOR EFFICIENT INFORMATION COMPUTING IN SOCIAL NETWORKS

Dr. (Mrs.) Pushpa R. Suri[1], Harmunish Taneja[2]

[1] Department of Computer Sc. and Appl., Kurukshetra University, Kurukshetra, India

pushpa.suri@yahoo.com

[2] Department of Information Technology, M.M. University, Mullana, India

harmunish.taneja@gmail.com

## ABSTRACT:

*Social networks have ensured the expanding disproportion between the face of WWW stored traditionally in search engine repositories and the actual ever changing face of Web. Exponential growth of web users and the ease with which they can upload contents on web highlights the need of content controls on material published on the web. As definition of search is changing, socially-enhanced interactive search methodologies are the need of the hour. Ranking is pivotal for efficient web search as the search performance mainly depends upon the ranking results. In this paper new integrated ranking model based on fused rank of web object based on popularity factor earned over only valid interlinks from multiple social forums is proposed. This model identifies relationships between web objects in separate social networks based on the object inheritance graph. Experimental study indicates the effectiveness of proposed Fusion based ranking algorithm in terms of better search results.*

## 1. INTRODUCTION

Searching from billions of objects, makes it challenging to harness single right result. Also, ranking needs to be redefined beyond the boundaries of query driven search in today's scenario. Traditionally the entities in each search results are similar in sense they are web pages. With the proliferation of social sites there is a big bang explosion of the web users that are not just passive. The trade-off between the systems costs of extracting a single right web object and a set of potential results continues to be major design issue. Enhanced ranking needs to be prepared for selecting and merging heterogeneous results such as videos, images, news and local business listings and not just text features. Web expansion has transformed the "just enter in Google" search approach to the "likes" button of Facebook that allows to submit personalized opinion on absolutely any object of interest on web. The "likes" play significant role in increasing or decreasing the popularity of the web object contained in social network. Social networks have paced the wave for involving users at all level of search and creating a raised level of interest in manual indexing [1]. Consequently its impact on the ranking method is imperative. All search engines use some kind of web object link-based ranking in their ranking algorithms. This has been the result of the success of Google, and its PageRank link algorithm [2]. Traditionally from small amount of labeled data to more recently, machine learning technology are exploited to create ranking function. But existing technologies are inadequate for search applications offered by various modern social networks. A new learning framework that could rank related web objects that may be heterogeneous from multiple distinct and or similar social forums is desired. Vertical search engines refer to the search services restricted by specific information, such as music, image, video and academic search [3]. Popular vertical search engines like Academic Search [4], Google Scholar[5], Froogle [6], Movie Search [7], Image Search [8][9], Video Search [7], News Search [10] and many more have become increasingly effective in serving users with specific needs. But social networks reflect real world data sets as heterogeneous, semi-structured and multi-relational. Ranking function is most critical part of the search engine with two key factors viz. relevance and quality [11].

The social network users are the volunteers to actually rate the music which is a development to be noted for enhancing the traditional ranking methodology. If search results for query based on a song are coming from different social networks, then it is not an easy task to rank them because ranking of song in one social network may be higher and in the second it may be less. Motivated by such observation, the proposed framework i.e. FBR (Fusion based Rank) for Social networks fuse the ranking scores of searched web objects (music) from heterogeneous social networks by identifying the relation link between web objects from different social forums. We propose an integrated ranking method in heterogeneous social network by recognizing relatedness among different web objects that could be reflected in the ranking particularly a more famous a singer is, the more likely the songs will receive majority likes. Consequently more famous songs lead to more popularity of singers than those that are less cited. The paper is organised as follows. Section 2 outlines the applications of web information computing. Section 3 elaborates the related work highlighting the challenges and choices faced in ranking web objects from multiple social networks and presents the attribute based comparison of popular social networking sites. Section 4 and 5 extends the concept of ranking and the social networks and global ranking over web. Section 6 elaborates the framework for fusion based web object ranking in social networks. The Ranking algorithm is proposed in Section 7. Section 8 describes the experimental study. The results on the effectiveness of the proposed Fusion Based Rank framework are discussed in Section 9 and finally the paper is concluded in Section 10.

## 2. APPLICATIONS OF WEB INFORMATION COMPUTING

One of the major inventions of the 1990s is the WWW. Since its advent in 1991, the web has evolved as global interconnection of individual networks operated and used by both public and private sectors. Its origin can be traced back when Internet was the tool to interconnect laboratories busy in government research.

WWW has been exponentially expanding reaching every remote corner of the earth. It has revolutionized the availability of electronically accessible information. Historically the information computing methodologies implemented were sufficient to tackle only 623 websites on the internet (1993), than today, when there are over 165 million websites on the internet. The graph quoted in Figure 1 [12] further approves that enhancing the rudiments of traditional information computing over web is the need of the hour.

Web information computing has spectrum of applications from academic research to commercial applications. The enormous increase in the quantity of online text available and the demand for access to different types of information by diverse users have, however, led to transformed applications of web information computing in a wide range that go beyond simple web page retrieval, such as question answering, web object detection and tracking, summarization, multimedia retrieval, biological and chemical informatics, searched results mining, software engineering and genomics [13]. The applications of web information computing also include the content search in the digital libraries, media search like image, music, news, blog and video retrieval. Web information computing provides a flexible and open platform for electronic commerce and other peer to peer approaches, searching on the desktop, enterprise, federated, mobile, and social search for varied web users.

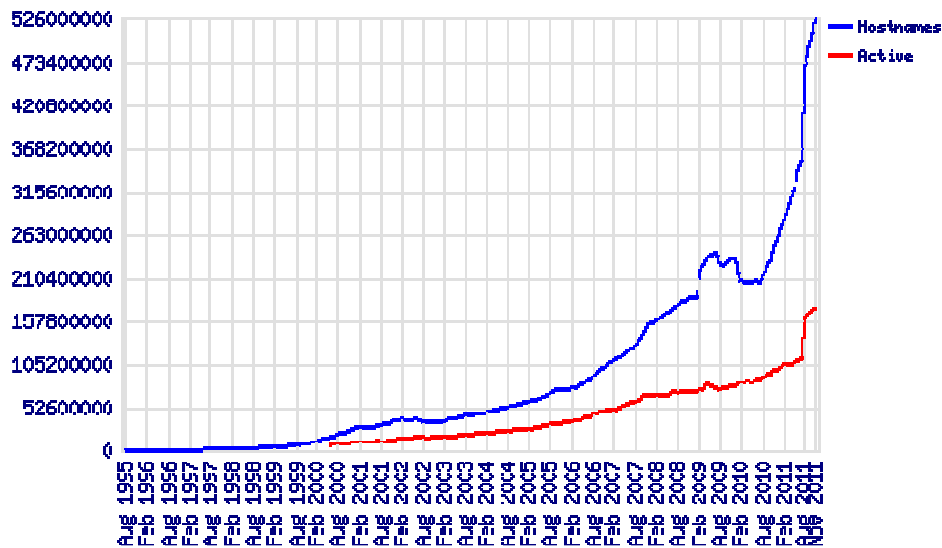

Figure1: Total Sites across all Domains Aug. 1995 - Nov. 2011 [12]

Domain specific applications of information computing include expert search, geographic information computing, and web information computing in specific domains and other vertical searches.

## 3. BACKGROUND

Traditionally search engines have only had access to three major types of data describing web pages, i.e., web page content, link structure and query or click through log data. A not so new entry to this family of document types is the user generated content describing the pages directly is less explored [14]. Networked web pages and their ranking has been noticed at length like PageRank [13], HITS [15]. The theory of PageRank is very simple, i.e, "good pages reference good pages" and are labelled with have higher PageRank. But ranking social network web objects, on the other hand, is employed for exploring links between similar web object from heterogeneous origins. Object-level ranking has been well studied for vertical search engines. But algorithms evolved from PageRank [13], PopRank [3] and LinkFusion [16] proposed to rank objects coming from multiple communities suggests that the ranking is restricted by the unavailability of well-defined graphs of heterogeneous data in some kinds of vertical domains in which objects.

The challenge of integrating ranks of web objects from multiple heterogeneous social networks lies in their distinct ranking systems. The ranking interval or scale including the minimal and maximal ratings for each web object is varied from 5 points to 10 points rating scale. Another important issue is the rating criteria of different social forums. Also there may be ambiguity in the quality of the same score [17]. Future of WWW is where individuals are linked together in addition to web pages. The rise of social networks and its hold over common man interest from academician to a naive web surfer indicates the inefficiency of traditional search theories in general.

Table 1 presents the comparative elaboration of the top fifteen most popular social networking sites with attributes like monthly visitors and various ranks (as on July 13, 2011) *eBizMBA Rank* [18], *Alexa* Global Traffic Rank [19] and U.S. Traffic Rank from both *Compete* [20] and *Quantcast* [21] that are constantly updated at each website's.

Table 1. Comparison of the Top 15 Most Popular Social Networking Sites

| Name | Logo | Estimated Unique Monthly Visitors | eBizMBA Rank | Compete Rank | Quantcast Rank | Alexa Rank |
|---|---|---|---|---|---|---|
| Facebook | 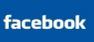 | 700000000 | 2 | 3 | 2 | 2 |
| Twitter | 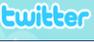 | 200000000 | 15 | 30 | 5 | 9 |
| Linkedin | 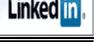 | 100000000 | 33 | 57 | 26 | 17 |
| mySpace | 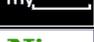 | 80500000 | 50 | 26 | 44 | 79 |
| Ning | 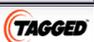 | 60000000 | 143 | 180 | 120 | 128 |
| Tagged | 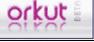 | 25000000 | 255 | 382 | 151 | 141 |
| Orkut | 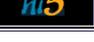 | 15500000 | 401 | 570 | 540 | 93 |
| hi5 | 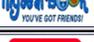 | 11500000 | 479 | 983 | 392 | 62 |
| myyearbook | 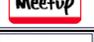 | 7450000 | 617 | 522 | 293 | 1036 |
| Meetup | 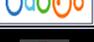 | 7200000 | 635 | 644 | 732 | 528 |
| Badoo | 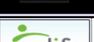 | 7100000 | 653 | 1346 | 489 | 125 |
| bebo | 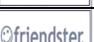 | 7000000 | 655 | 944 | 434 | 588 |
| mylife | 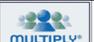 | 5400000 | 865 | 118 | 688 | 1789 |
| friendster | 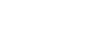 | 4900000 | 955 | 1920 | 643 | 301 |
| Multiply | 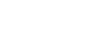 | 4300000 | 1136 | 2446 | 677 | 285 |

## 4. Ranking and Social Networks

Ranking as the name suggests deals with ordering the search results of a query according to a certain measure of relevance to the user's query. Information computing in general has ranking techniques that greatly simplify the user interaction with large search results in the predefined relevance. Ranking estimates the quality of a set of results retrieved by a search engine [22]. Traditional information computing advocates the use of major boolean, probabilistic, or vector-space models. Modern information computing promotes the link structure of the web as links supports positive critical review of a web page's content which originates from outside of the control of the page's author. This makes the information extracted from informative links less vulnerable to manipulative techniques such as spamming [13], [15], [23], [24] still many ranking issues remain to be explored.

Complimentary to the emergent face of internet, the web user has also shown an embryonic growth not only in number but also in diversity that has spurred research challenges for matching users with interesting web content. Particularly, social networks offers many indications reflecting interests of a user such as the links in the network, user profiles, likes buttons and text in user-generated content. Such heterogeneous information which is mostly unstructured from varied sources lacks to model relevance that further mount to the challenges of the web information processing. Social networking services such as Facebook, Youtube and Twitter are as important as oxygen for many making it an indispensable component of the web.

## 5. Global Ranking of Web pages

The Random Surfer model [25] and the page rank based selection model [26] are described as two major models [27]. Page rank based selection model tries to capture the effect that the search engines have on the growth of the web by adding new links according to page rank. The page rank algorithm is used in the Google search engine [28] for ranking search results.

Google is a prototype of a large-scale search engine that makes heavy use of the structure present in hypertext [13]. Google is designed to crawl and index the web efficiently and produce much more satisfying search results than existing systems. Link Analysis Ranking [29] emphasize that hyperlink structures are used to determine the relative authority of a web page and produce improved algorithms for the ranking of search results. The prototype with a full text and hyperlink database of web pages is available at [30]. In the current era there is much concern in using random graph models for the web [31].

PageRank is a link analysis algorithm used by the Google Internet search engine that assigns a numerical weighting to each element of a hyperlinked set of documents, such as the WWW, with the purpose of "measuring" its relative importance within the set. Google is designed to be a scalable search engine with primary goal to provide high quality search results over a rapidly growing WWW [32]. Link-based ranking algorithms rank web pages by using the dominant eigenvector of certain matrices like the co-citation matrix or its variations [13]. Distributed page ranking on top of structured peer-to-peer networks is needed because the size of the web grows at a remarkable speed and centralized page ranking is not scalable [33]. Page ranking can be propagation rates depending on the types of the links and user's specific set of interests [34]. Page filtering can be decided based on link types combined with some other information relevant to links. For ranking, a profile containing a set of ranking rules to be followed in the task can be specified to reflect user's specific interests [35]. Similarities of contents between hyperlinked pages are useful to produce a better global ranking of web pages [36].

## 6. FBR (FUSION BASED RANK): FRAMEWORK

Manually the ranking criterion can be standardized by score normalization and transformation if the search involves same song or singer [37]. This section presents an integrated ranking framework FBR to analyse inter- and intra-type links and to sort web objects in different social networks at the same time. The conceptual scheme of proposed integrated ranking framework is shown in Figure 2.

Two kinds of links among web objects are defined: intra-type links: which represent the relationship among web objects within a social network, and inter-type links: which represent the relationship among web objects between different social networks.

FBR exploits the two layer hierarchical approach as shown in Figure 2 for identifying the inheritance and aggregation relationships between classes by creating super classes that may encapsulate the services or features common to other classes.

There are k graphs $(S_1, S_2,......,S_k)$ where each $S_i = (O_{in}, E_i)$ is the un-weighted undirected graph (social network) of web objects $O_{in}$ with links as $E_i$. The classes may belong to different social networks $S_1, S_2,...S_k$. Social Network $S_i$ has a collection of related web objects $(S_i: O_{i1}, O_{i2}, O_{i3}.....O_{in})$ where i varies from 1 to k and k is the social networks integrated. If the web objects of different social networks share such a relationship as "has" or "contains" between them, an aggregation link is identified which is further implemented for rank computation.

The rank computed by each social forum is labelled as the partial rank. The proposed framework add a popularity factor based on each interlink of the web object pointing to the queried object based on object inheritance graph. Different types of relationship are given different popularity factor.

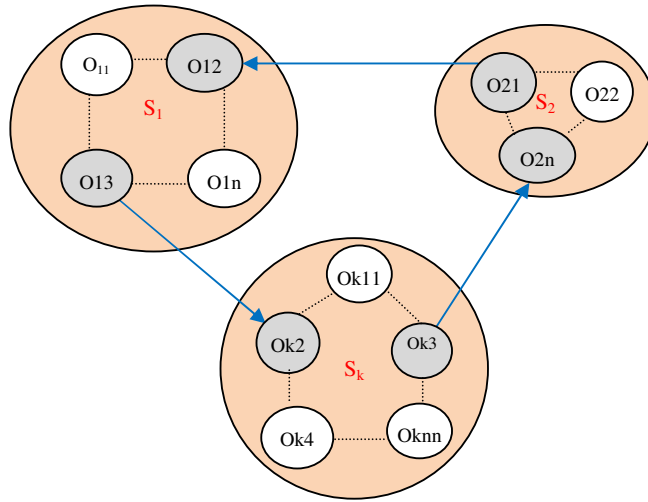

Figure 2. Integrated social networks for enhanced Fusion based ranks: FBR

## 6.1 COMPONENTS OF FUSION BASED RANKING FRAMEWORK.

The proposed model calculates the rank for web objects with the activity flow of the FBR method as shown in Figure 3. Essential components of FBR framework are Query Handler, Link Administrator and Relationship score Manager and Fused Global Rank. Figure 3 elaborates the basic configuration and working of the proposed model. The main components of are as under:

- **Query Handler:** When the users request based on query terms, the invocation goes to the Query Handler. It extracts one or more web objects from the entered query and filters similar requests as single and forwards it to various social networks.

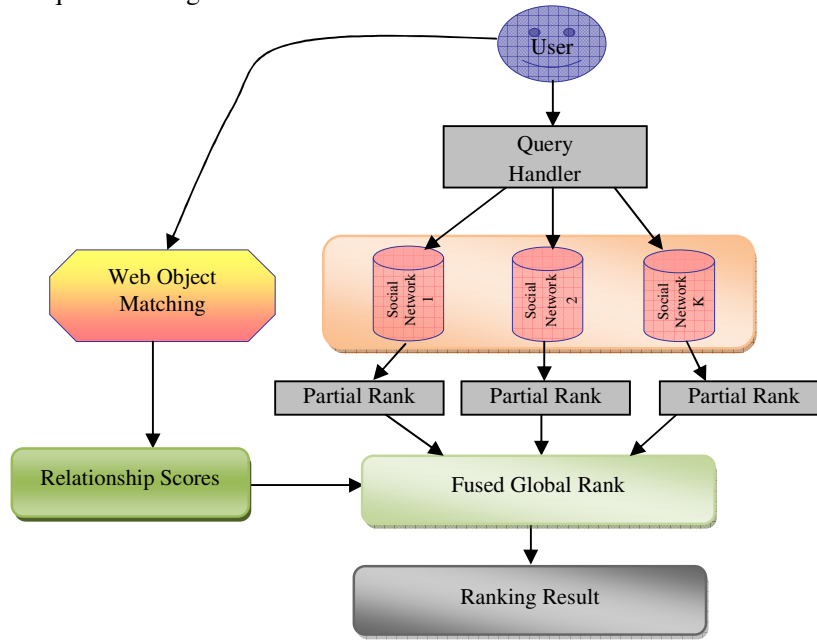

Figure 3. Activity flow of FBR Method

- **Web Object Matching**: All the related web objects in various social networks contribute to the final rank of the search results. This module filters the related objects in social networks under study. It searches the desired web objects from the various repositories over web. It stores the domain specific feature set as (F1,F2, . . . , Fm) where m is the total features under consideration. It stores inheritance network graph corresponding to the specific feature set reflecting the relatedness among the web objects from various social networks sharing attributes inherited from its super classes. The retrieved web objects defined as set of features is matched for their association with the generated inheritance graph. The relevant interlinks are further handled and labeled as type *valid* and forwarded to the fused global rank or if found not valid are discarded.

- **Relationship score**: Relationships among objects contribute to their popularity factor and this module computes the score of searched web objects in terms of its valid interlinks over other social networks.

- **Fused Global Rank**: This module computes the final Global rank indicating the actual level of relevancy of the search results.

## 7. PROPOSED RANKING ALGORITHM: FUSION_BASED_RANK

The Algorithm for FBR method derived from proposed framework is as below:

| Algorithm: Fusion_Based_Rank | |
|---|---|
| Step 1 | The Users / Clients request for desired web information. |
| Step 2 | Query handler Processes the Clients requests and filters similar requests as single. |
| Step 3 | Web objects Partial ranks are computed by performing link analysis equipped with user's feedback as manual indexes. |
| Step 4 | Partial ranks are furnished for each social network $S_i$ by exploiting Intra-type links. |
| Step 5 | Web Object Matching  module generates the relationship function $Rel_i$ for $i^{th}$ retrieved web object as<br><br>$Rel_i = F((f1, f2, \ldots \ldots, fn), \text{€})$<br><br>where web object feature set (f1,f2, . . . . . ,fn) are the features matched with the domain feature set (F1,F2, . . . . . , Fn) and a constant threshold € is the minimum number of features of the web object *i to be* matched with domain feature set for it to be a member of the inheritance network graph of that domain. |
| Step 6 | IF n (number of features of the web object *i* matched with domain feature set) > €<br><br>    THEN            $Rel_i$ = 1<br><br>    ELSE                  = 0 |
| Step 7 | IF  $Rel_i$ = 0<br><br>THEN $i^{th}$ web object interlink is not valid<br><br> ELSE  $i^{th}$ web object interlink is valid and contribute to the popularity factor of the web object. |
| Step 8 | The popularity factor is combined with the partial ranks to estimate a global rank for the searched web object. |

## 8. EXPERIMENTAL STUDY

There exist a number of redundant popular songs between any two social forums, because music web site may submit song to many social forums in order to obtain the feedback or comments. The proposed ranking method FBR emphasizes that although the ranks of duplicate songs in different social forums should be almost equal but practically this is not the case. The rank variation is ruled by the interest of the people connected to the social network. In the proposed work, Object orientation concepts help identifies the relationships of the searched web object within and across social networks. Each social network is viewed as a sub graph where each web object and its relationship with other web objects is reflected in the form of multiple level inheritance.

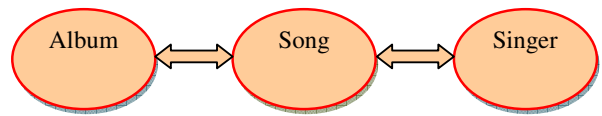

Figure 4. Music Object relationship graph

The attribute value of web objects is obtained by iteratively calculating the periodic user's feedback. It is clear that the more popular the web object is, the more likely the user is interested in it. For ranking the most unique feature of the proposed approach is the web object graph with heterogeneous links e.g. a music object song may be searched by a set of other music objects that may be sung by a set of singers object and included in other web object album as shown in Figure 4.

## 8.1 Datasets

Our study requires two sets of data: web objects and popular social networks. We explored various options for datasets, and eventually decided not to use data from popular social network services, such as Orkut, Facebook as the datasets in these sites are rich on social networks, but poor on web documents. We used data from YouTube which offers both rich web document data and extensive social network information. We first started with ten internet users randomly chosen with high degree and different interests, and obtained their friends and music interests. Then we used these friends as the new players and fetched the friends and music metadata from these. This process was iterative and stopped until no more options were available or the number of retrieved web users exceeded a pre-defined threshold.

## 8.2 Web Object Data

In the YouTube music dataset, each web object contains rich metadata such as album title, genre, tags, popular artists, description, uploader, rating, etc. These metadata were downloaded with our crawler and stored locally. The music is labelled with 10 main categories (Pop, Rap, Rock, Jazz, Today's hit, Reggae, Country, International, Comedy, and Latin etc.) as provided by YouTube. In this experiment, we defined a user's interest with one of these music categories. If the majority of a user's videos fall into one category, we regard that web object with higher relatedness score. With the metadata of music web objects, the indexes for music documents are built. The values in the fields: title, tags, genre and description were parsed into terms to create an inverted index using, in which each term in a specific field points to a collection of music documents. These indexes can be easily used to generate term document relatedness scores.

## 8.3 Social Network

In YouTube, each user has attributes such as name, gender, location, date of birth, friends, music uploaded, and about-me. Social networks can then be constructed based on the information obtained. An inter-type link was created between two web objects if they exist in similar context. In this experiment, we used two fully connected social networks based on the downloaded music data:

- $S_1$: Social Network 1: a larger social network that consists of 18,576 different registered users and 39,271 music uploaded by users;

- S$_2$: Social Network 2: a smaller network that has 2,464 users and 7,978 music uploaded by users.

## 8.4 Evaluation

To examine the effectiveness of the proposed Fusion based ranking algorithm, we compared following two algorithms:

**Baseline:** tf-idf [11] was used as the base line method.

**Fusion based Rank (FBR):** *FBR*(Q,O,Rel) = *f*(S$_i$ Rank(Q, Web object), Rel (Rel Score) )
where Q is the query, Rel score is the similarity value between web objects in social networks. tf-idf values are normalized.

In each category, we choose ten queries that are related to the music category but at the same time could also refer to things that do not belong to the category. Queries in the music category were about album titles, singer etc. The top twenty returned music of each query were mixed and presented to two web surfers for independent evaluation. They evaluated each returned result based on to what extent the returned music was relevant to a query as well as to what extent the music was related to the user's interest. Table 2 shows the scores used to rate the relevance of a return. The highest score of a return is 5 and the lowest score is 0. After a pilot experiment, the inter-rater reliability among two raters was 70%, indicating that they reached a reasonable agreement about the relevance criteria [38].

Table2.  Evaluating Content Relevance

|        | High | Medium | Low |
|--------|------|--------|-----|
| High   | 5    | 3      | 0   |
| Medium | 3    | 2      | 0   |
| Low    | 0    | 0      | 0   |

We used *Normalized Discounted Cumulative Gain (NDCG)* [39] metric to evaluate the ranking algorithms. This method measures the usefulness of the ranking result based on the relationship between the relevance scale of web objects and their position in the ranking. The premise of DCG is that highly relevant documents are more useful when they have higher ranks in the result list. The NDCG at position *k* is given by:

$$NDCG_K = \frac{DCG^R{}_K}{DCG^T{}_K}$$

$$DCG^X{}_K = \mathrm{Re}\,l_1(X) + \Sigma \frac{\mathrm{Re}\,l_i(X)}{\log_i}$$

where *i  goes from 2 to k*

*Rel$_i$(X)* shows the level of relevancy for the result at position *i* in rank X.

*DCG$^T{}_K$* is the value for the optimal rank at position *k*.

# 9. RESULTS AND DISCUSSIONS

We evaluated the performance of two different ranking approaches. The NDCG values were averaged over all queries without considering the differences in the size of social networks and the degrees. As shown, FBR method performed better than the baseline algorithm. Figure 5 shows that, for the first 20 search results, the NDCG values of FBR are higher than that of the baseline algorithm.

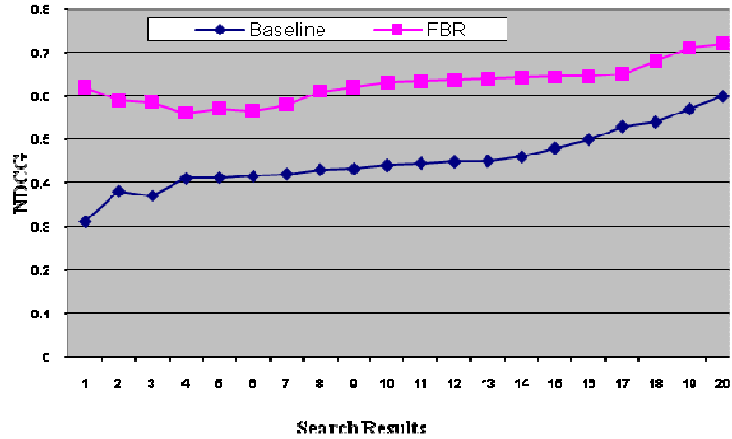

Figure 5. NDCG: Fusion Based Rank Vs Baseline

The performance of these two algorithms is compared in two social networks with different size. To make the results more comparable, we only used searchers for the music category in both networks. The NDCG results from two social networks are shown in Figure 6.

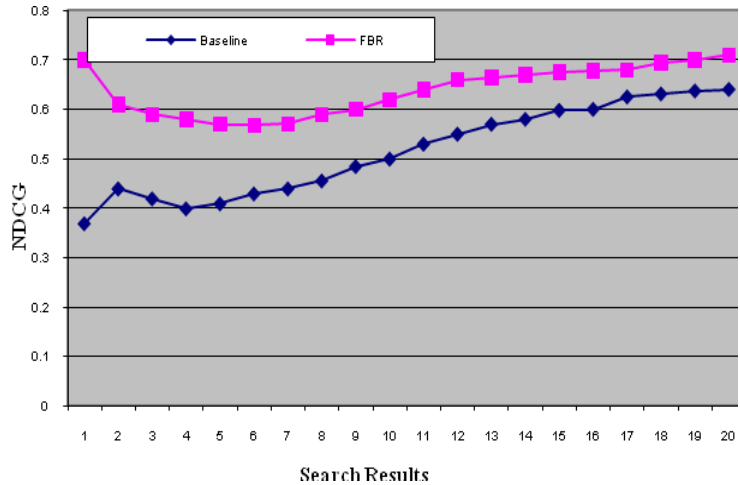

Figure 6. NDCG: Fusion Based Rank Vs Baseline for Larger Social Network 1

The NDCG scores are averaged over all queries from searchers of music for the two networks. As shown, for the larger social network 1, the FBR method performs better than the baseline algorithm. But the difference in the smaller social network 2 is minimal as shown in Figure 7.

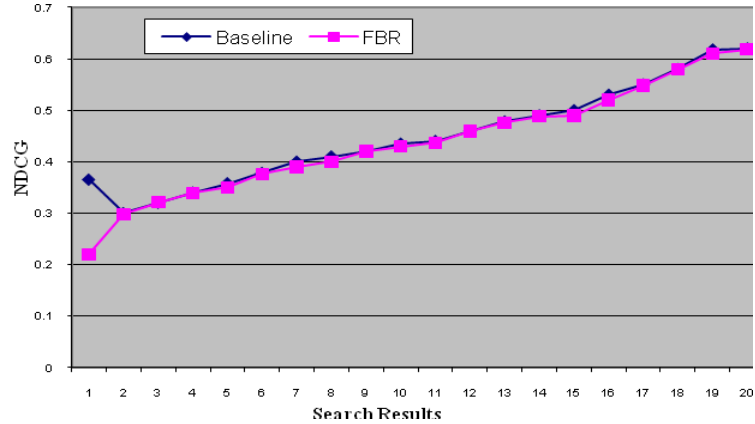

Figure 7. NDCG: Fusion Based Rank Vs Baseline for Small Social Network 2

## 10. CONCLUSION

Social networks facilitate the formation of user generated content in diverse formats. Such networks smooth not only the classification user oriented web objects but also social indexing. Users send comments and feedbacks, likes or dislikes resources, upload and share multimedia contents, communicate online with social contacts, pour in wiki-style knowledge bases, maintain personal bookmarks.

In this paper, the web object ranking problem in case of social networks where traditional ranking algorithms are no longer valid is explored. The fusion based rank (FBR) method integrates many social networks and identifies the relationship graph considering the valid interlinks among the related or duplicate web objects. The results of our evaluation studies indicate that overall, the FBR framework can return better search results than the traditional tf-idf ranking algorithm in terms of relevance, the ranking effectiveness of returned search results as it considers the global information of a social network. But its performance is appreciable only as the network size reaches a certain magnitude. FBR method is more effective than the baseline approach as it considers the global information of social networks. FBR method benefits more from large social networks than from small networks.

**Dr. Pushpa R. Suri** received her Ph.D. Degree from Kurukshetra University, Kurukshetra. She is Associate Professor in the Department of Computer Science and Applications, Kurukshetra University, Kurukshetra (KUK), Haryana, India. Her publications are in International and National Journals and Conferences of repute. Her teaching and research activities include Discrete Mathematical Structure, Data Structures, Information Computing and Database Systems.

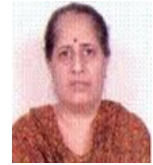

**Harmunish Taneja** is pursuing Ph.D. (Computer Science) from Kurukshetra University, Kurukshetra. He has received his M.Phil. (Computer Science) degree from Algappa University, Tamil Nadu, India and MCA from Guru Jambeshwar University of Science and Technology, Hissar, Haryana, India. Currently he is working as Assistant Professor in the Department of Information Technology at M.M. Engineering College, M.M. University, Mullana, Haryana, India. He has published 20 papers in International / National Conferences and Seminars. His teaching and research areas include Database systems, Web Information Retrieval, and Object Oriented Information Computing.

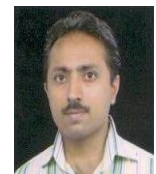